\documentclass[aps,prl,twocolumn,showpacs]{revtex4}
\usepackage{bm}

\usepackage{amsmath}
\usepackage{amssymb}
\usepackage{graphicx}
\usepackage{psfrag}

\newcommand{\beq}{\begin{equation}}
\newcommand{\eeq}{\end{equation}}
\newcommand{\beqar}{\begin{eqnarray*}}
\newcommand{\eeqar}{\end{eqnarray*}}

\newcommand{\ua}{\uparrow}
\newcommand{\da}{\downarrow}

\newcommand{\dm}{\diamond}

\newcommand{\Ktxt}{{\text K}}

\newcommand{\Stxt}{{\text S}}
\newcommand{\Ttxt}{{\text T}}
\newcommand{\Utxt}{{\text U}}

\newcommand{\Tc}{\mathcal{T}}

\newcommand{\rtarr}{\rightarrow}

\newcommand{\nb}{{\bf n}}

\newcommand{\s}{{\bf s}}

\newcommand{\rb}{{\bf r}}

\newcommand{\Bt}{{\tilde{B}}}

\newcommand{\dg}{\dagger}

\newcommand{\ran}{\rangle}

\newcommand{\al}{\alpha}

\newcommand{\De}{\Delta}

\newcommand{\vphi}{\varphi}

\newcommand{\e}{\epsilon}

\newcommand{\tr}{\text{tr}}

\newcommand{\Ec}{{\cal{E}}}

\begin{document}

\title{Canted antiferromagnetic phase of the $\nu=0$ quantum Hall state in bilayer graphene}

\author{Maxim Kharitonov}

\address{
Center for Materials Theory,
Rutgers University, Piscataway, NJ 08854, USA
}
\date{\today}
\begin{abstract}

Motivated to understand the nature of the strongly insulating $\nu=0$ quantum Hall state in bilayer graphene,
we develop the theory of the state in the framework of quantum Hall ferromagnetism.
The generic phase diagram,
obtained in the presence of the isospin anisotropy,
perpendicular electric field, and Zeeman effect,
consists of the spin-polarized ferromagnetic (F),
canted antiferromagnetic (CAF),
and partially (PLP) and fully (FLP) layer-polarized phases.
We address the edge transport properties of the phases.
Comparing our findings with the recent data on suspended dual-gated devices, we conclude that the insulating $\nu=0$ state
realized in bilayer graphene at lower electric field is the CAF phase.
We also predict a continuous and a sharp insulator-metal phase transition
upon tilting the magnetic field from the insulating CAF and FLP phases, respectively,
to the F phase with metallic edge conductance $2e^2/h$, which could be within the reach of available fields
and could allow one to identify and distinguish the phases experimentally.

\end{abstract}
\maketitle

\begin{figure}
\centerline{\includegraphics[width=.28\textwidth]{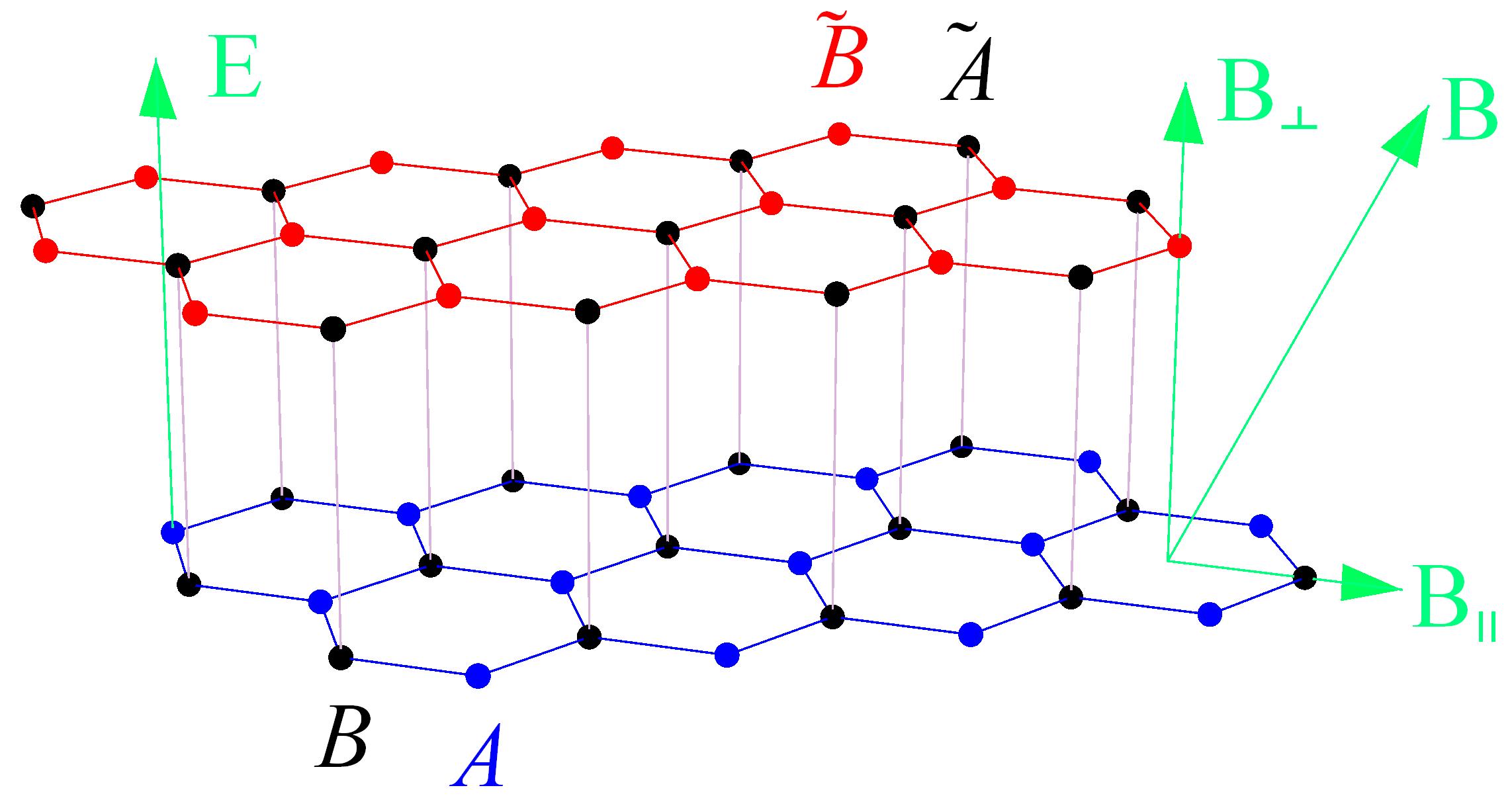}\hspace{2mm}
\includegraphics[width=.15\textwidth]{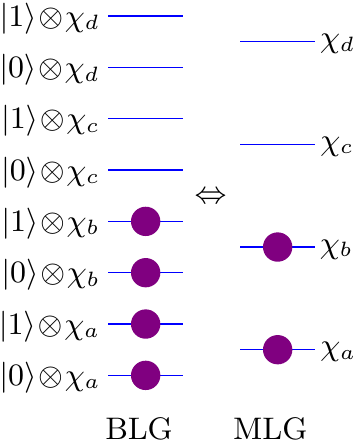}}
\caption{(color online). (left)
Lattice structure of BLG.
(right) Occupation of the $01\otimes KK'\otimes s$ space of each orbital by four electrons in the $\nu=0$ QHFM state in BLG.
The energy of the $KK'\otimes s$-symmetric interactions is minimized by forming 01-pseudospin-singlet pairs~\cite{Barlas,APS}.
Correspondence between the $\nu=0$ QHFM states in BLG and MLG is shown.
}
\label{fig:basic}
\end{figure}

{\em Introduction.}
One of the most intriguing questions in today's graphene research concerns the nature of the
strongly insulating $\nu=0$ quantum Hall (QH) state
[with half-filled zero-energy Landau level ($\e=0$ LL)],
observed in both monolayer (MLG)~\cite{MLG} and bilayer (BLG)~\cite{HarvardBLG,ColumbiaBLG,h,SchweizBLG,Velasco} graphene
with two-terminal conductance of the highest quality samples $G \lesssim 10^{-5} e^2/h$.
While the basic theoretical framework of the interaction-induced $\nu=0$ state
-- the concept of generalized quantum Hall ferromagnetism (QHFMism)~\cite{QHEbook} --
is well-established~\cite{NM,YDM,ee,JM,eph,Barlas,APS,GorbarBLG,NL2,MKmono},
it is unambiguously identifying the particular order of the $\nu=0$ QHFM
that presents a challenge.
Given the rich phase diagram of the $\nu=0$ QHFM in MLG~\cite{ee,JM,eph,MKmono} (and as we show here, in BLG)
and the fact that all phases but the spin-polarized one~\cite{Abanin,FB}
are expected to be fully insulating~\cite{JM,Gusyninedge,MKprep},
achieving this goal requires a more detailed both theoretical and experimental  analysis.

On the experimental side, in BLG,
a crucial step in this direction  was recently made
in dual-gated suspended devices~\cite{h,Velasco},
where application of perpendicular electric field $E$ offers a unique
possibility to manipulate the layer ``isospin''.
At perpendicular magnetic fields $B_\perp \gtrsim 1 \Ttxt$,
upon applying the electric field,
a phase transition
to yet another insulating QH state with $G \ll e^2/h$ was observed,
which can be readily identified as the valley=layer-polarized phase of the $\nu=0$ QHFM.
This transition was characterized by a spike in conductance with maximum $G \sim e^2/h$
at the critical field $E^* \approx 11 B_\perp[\Ttxt] \text{meV}/\text{nm} $
and was observed for both polarities of the electric field.

Motivated by this result, in this Letter
we develop the theory of the $\nu=0$ QHFM in BLG.
We obtain a generic phase diagram
in the presence of the isospin anisotropy
of electron-electron (e-e)~\cite{ee,JM,MKmono} and electron-phonon (e-ph)~\cite{eph,MKmono} interactions,
electric field, and Zeeman effect.
We address the edge transport properties of the phases.
Comparing our findings with the data of Refs.~\onlinecite{h,Velasco},
we arrive at the conclusion that
the insulating $\nu=0$ QH state realized
in BLG at lower electric field~\cite{HarvardBLG,ColumbiaBLG,h,SchweizBLG,Velasco} is the canted antiferromagnetic phase
of the $\nu=0$ QHFM:
the very existence of the phase transitions with applied electric field provides
sufficient information for that.
We also predict that experiments in the tilted magnetic field could
verify this conclusion and
allow for observation of new phase transitions.

{\em $\nu=0$ QHFM in BLG.} Our analysis follows closely that for MLG~\cite{MKmono},
and details will be presented elsewhere~\cite{MKprep}.
The $\e=0$ LL in BLG, located at the charge neutrality point, possesses
very peculiar properties~\cite{MF}.
First, analogous to the case in MLG, in each valley, $K$ or $K'$, its wave functions
reside on only one sublattice, $\Bt$ or $A$, of the low-energy two-band model~\cite{MF} and hence in either one of the layers
(Fig.~\ref{fig:basic}).
This makes not only the $A\Bt$ sublattice and layer, but also the valley degree of freedom
equivalent, $K \leftrightarrow \Bt$, $K' \leftrightarrow A$ (referred to as $KK'$ ``isospin'' here).
Second,
both
$|0\ran$ and $|1\ran$
magnetic oscillator states belong to the $\e=0$ LL,
which results in its unique extra twofold orbital degeneracy
(this subspace is referred to as $01$ ``pseudospin'' here).
Each orbital of the $\e=0$ LL is thus
(approximately) eightfold degenerate in the $01 \otimes KK' \otimes s$ pseudospin-isospin-spin  space.

According to the general theory of QHFMism~\cite{QHEbook}, at integer filling factors $\nu$,
Coulomb interactions result in spontaneous ordering of
discrete degrees of freedom (spin, valley, etc), favoring the many-body states,
in which each orbital is occupied by electrons in exactly the same way.
At the $\e=0$ LL in BLG,
due to the difference in wave functions of the $|0\ran$ and $|1\ran$ states,
interactions possess an intrinsic anisotropy in the 01-pseudospin space~\cite{Barlas,APS}.
As demonstrated in Refs.~\cite{Barlas,APS},
at $\nu=0$ the energy minimum of the $KK'\otimes s$-symmetric interactions
is delivered by those QHFM states, in which four electrons per orbital occupy
the states $|0\ran \otimes \chi_a$, $|1\ran \otimes \chi_a$,  $|0\ran \otimes \chi_b$, $|1\ran \otimes \chi_b$
with arbitrary orthogonal spinors $\chi_{a,b}$ in the $KK'\otimes s$ space,
i.e., form two pseudospin-singlet pairs (Fig.~\ref{fig:basic}).
Ordering of the remaining
isospin-spin degrees of freedom
is governed by (weaker) mechanisms of the $KK' \otimes s$-symmetry breaking.
Following Ref.~\cite{MKmono}, the energy (per orbital per electron in a pair)
\begin{eqnarray}
    \Ec(P) &=& \Ec_\dm(P) + \Ec_V(P) + \Ec_Z(P),
\label{eq:Ec}\\
    \Ec_\dm(P) &=& \frac{1}{2}\sum\nolimits_{\al} u_\al \{ \tr^2 [\Tc_\al P ] - \tr [\Tc_\al P \Tc_\al P ] \},
\label{eq:Edm} \\
    \Ec_V(P) &=& -  \e_V \,\tr [ \Tc_z P ] ,\mbox{ }     \Ec_Z(P) = - \e_Z \,\tr [ S_z P ],
\label{eq:EVZ}
\end{eqnarray}
of these effects as a function of the order parameter matrix
$
    P = \chi_a \chi_a^\dg + \chi_b \chi_b^\dg
$
is obtained by calculating the expectation value of the microscopic Hamiltonian for BLG
with respect to the family of QHFM states.
Here,
$\al=x,y,z$, $\Tc_\al = \tau^{KK'}_\al \!\!\otimes \!\hat{1}^s$, $ S_z = \hat{1}^{KK'}\!\!\otimes\! \tau_z^s $, $\tau_\al$
are the Pauli matrices, and $\tr[\ldots]$ is the matrix trace.
The single-particle electric field [$\Ec_V(P)$] and Zeeman [$\Ec_Z(P)$] effects are characterized by the energies
$\e_V\approx  E a_z/2$~\cite{biasfield}, where $a_z \approx 3.5\AA$ is the
interlayer distance,
and $\e_Z = \mu_B B$,
where $B=\sqrt{B_\perp^2+B_\parallel^2}$
is the total magnetic field. The magnetic field $B$ has arbitrary direction relative to the sample (Fig.~\ref{fig:basic})
and the $z$ axis in spin space is chosen along it.
The many-body $KK'$-symmetry-breaking effects of e-e and e-ph interactions,
crucial in determining the preferred ground-state order,
give rise to the isospin anisotropy $\Ec_\dm(P)$.
Its generic form (\ref{eq:Edm}) is fully characterized by two signed $B_\perp$-dependent energies $u_\perp \equiv u_x=u_y$ and $u_z$.
The bare energies can roughly be estimated~\cite{ee,eph} as $|u_{\perp,z}^{(0)}| \sim e^2 a/l_B^2 \sim 1-10 B_\perp[\Ttxt] \Ktxt$,
where $a$ is some lattice spatial scale and $l_B =\sqrt{\hbar c/(e B_\perp)}$,
and can be further renormalized~\cite{MKmono}.

\begin{figure}
\includegraphics[width=.48\textwidth]{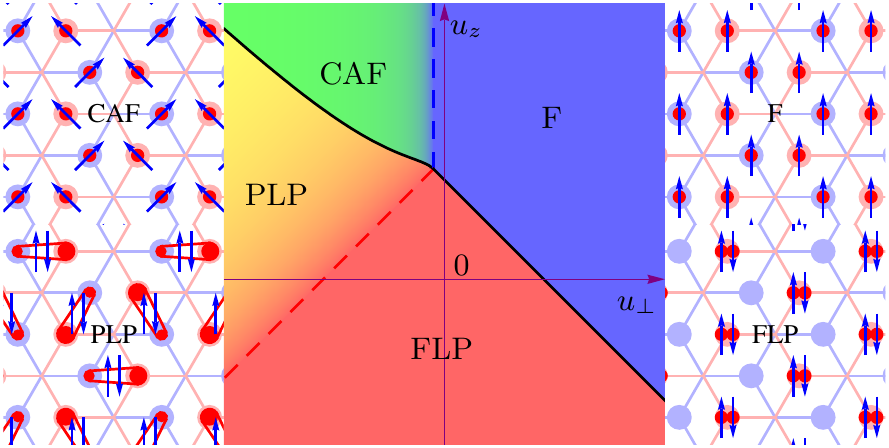}
\caption{(color online). Phase diagram of the $\nu=0$ QHFM in BLG in the space of isospin anisotropy energies $(u_\perp,u_z)$
at fixed electric $\e_V$ and Zeeman $\e_Z$ energies.
}
\label{fig:PD}
\end{figure}

{\em Phase diagram.}
The zero-temperature mean-field phase diagram is obtained
by minimizing the energy $\Ec(P)$~\cite{QFs}.
Remarkably, the theory of the $\nu=0$ QHFM in BLG described by Eqs.~(\ref{eq:Ec}), (\ref{eq:Edm}), and (\ref{eq:EVZ})
appears to be formally equivalent to that in MLG~\cite{MKmono},
upon identifying the pseudospin-singlet electron pairs in BLG
with single electrons in MLG (Fig.~\ref{fig:basic}):
$\{|0\ran, |1\ran \}\otimes \chi_{a,b}$
$\leftrightarrow \chi_{a,b}$.

In particular,
the phase diagram at zero electric field, $\e_V=0$,
is identical to that in MLG~\cite{MKmono}.
The anisotropy energy $\Ec_\dm(P)$ alone
is minimized by one of the following phases (Fig.~16 in Ref.~\cite{MKmono}):
spin-polarized isospin-singlet ferromagnetic (F, $\chi_a=|K\ran \otimes |\s\ran$, $\chi_b=|K'\ran \otimes |\s\ran$)
at $u_\perp>0$, $u_\perp+u_z>0$;
antiferromagnetic (AF, $\chi_a=|K\ran \otimes |\s\ran$, $\chi_b=|K'\ran \otimes |\! -\!\s\ran$),
with antiparallel spin polarizations $\pm \s$ of the layers, at $u_z>\!-u_\perp>0$;
and two isospin-polarized spin-singlet phases: fully  layer-polarized phase
[FLP, $\chi_a=|\pm \nb_z \ran \otimes |\ua \ran$, $\chi_b=|\pm \nb_z \ran \otimes |\da \ran$, $\nb_z=(0,0,1)$,
analogue of the charge-density-wave (CDW) phase in MLG] at $-u_\perp>|u_z|$
and, in the terminology of QH bilayers~\cite{QHEbook}, interlayer-coherent phase
[ILC, $\chi_a=|\nb_\perp \ran \otimes |\ua \ran$, $\chi_b=| \nb_\perp \ran \otimes |\da \ran$,
$\nb_\perp=(\cos \vphi_n,\sin \vphi_n,0)$, analogue of the Kekul\'{e} distortion (KD) phase in MLG] at $-u_z> |u_\perp|$.
Here and below, $\s$ and $\nb$ are the unit vectors defining
the spin and isospin polarizations of the states $|\s\ran$ and $|\nb\ran$, respectively;
$\pm \nb_z \leftrightarrow K,K'$ and  $\pm \s_z \leftrightarrow \ua,\da$.
The F and AF phases are $\Stxt\Utxt(2)$-spin-degenerate ($\s$),
and the ILC and FLP  phases are $\Utxt(1)$- and $\text{Z}_2$-isospin-degenerate ($\vphi_n$ and $\pm \nb_z$), respectively.
Including the Zeeman effect [minimization of $\Ec_\dm(P)+\Ec_Z(P)$, Fig.~18 in Ref.~\cite{MKmono}]
does not affect the spin-singlet ILC and FLP phases
but lifts the spin degeneracy of the F phase, $\s \rtarr \s_z=(0,0,1)$,
and transforms the AF phase to the $\Utxt(1)$-spin-degenerate ($\vphi_s$) canted antiferromagnetic phase
(CAF, $\chi_a = |K\ran \otimes |\s_a^*\ran$, $\chi_b = |K'\ran \otimes |\s_b^* \ran$), in which
the layers have noncollinear spin  polarizations
$
    \s_{a,b}^* = (\pm \sin \theta_s^* \cos \vphi_s, \pm \sin \theta_s^* \sin \vphi_s, \cos \theta_s^*)
$
with the optimal projection $s_z^* = \cos \theta^*_s =\e_Z/(2 |u_\perp|)$ on the total magnetic field.

Including the effect of electric field [minimization of $\Ec(P)$]
does not affect the F and CAF phases but lifts the isospin degeneracy
of the FLP phase, $\pm \nb_z \rtarr \nb_z$,
and transforms the ILC phase to the
partially layer-polarized phase
(PLP, $\chi_a = |\nb^* \ran \otimes |\ua \ran$, $\chi_b = |\nb^*\ran \otimes |\da\ran $),
in which  the valley=layer isospin
$
    \nb^*=(\sin \theta_n^* \cos \vphi_n, \sin\theta_n^* \sin \vphi_n, \cos \theta_n^*)
$
has the optimal value $n_z^* = \cos \theta_n^* = \e_V/(u_z+|u_\perp|)$
of the projection characterizing the degree of layer charge polarization.
As a result, in the presence of generic isospin anisotropy,  electric field, and the Zeeman effect,
the phase diagram of the $\nu=0$ QHFM in BLG
consists of the F, CAF, PLP, and FLP phases (Fig.~\ref{fig:PD}) 
with energies 
$\Ec^\text{F}=-2u_\perp-u_z-2\e_Z$, 
$\Ec^\text{CAF}=-u_z -\e_Z^2/(2|u_\perp|)$,
$\Ec^\text{PLP}=u_\perp-\e_V^2/(u_z+|u_\perp|)$, and 
$\Ec^\text{FLP}=u_z-2\e_V$.
The phase boundaries, obtained by comparing the energies,
are
$u_z-u_\perp = \e_V$ for PLP-FLP,
$u_\perp=-\e_Z/2$ for CAF-F,
$u_z+u_\perp=\e_V-\e_Z$ for F-FLP,
and
\beq
     u_\perp +  u_z = \e_V^2/(u_z-u_\perp) + \e_Z^2/(2 u_\perp)
\label{eq:CAFPLP}
\eeq
for CAF-PLP phase transitions, respectively.

\begin{figure}
\centerline{\includegraphics[width=.23\textwidth]{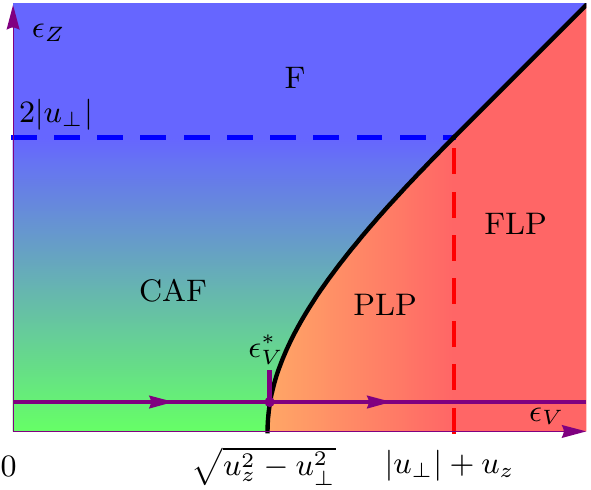}\hspace{2mm}
\includegraphics[width=.20\textwidth]{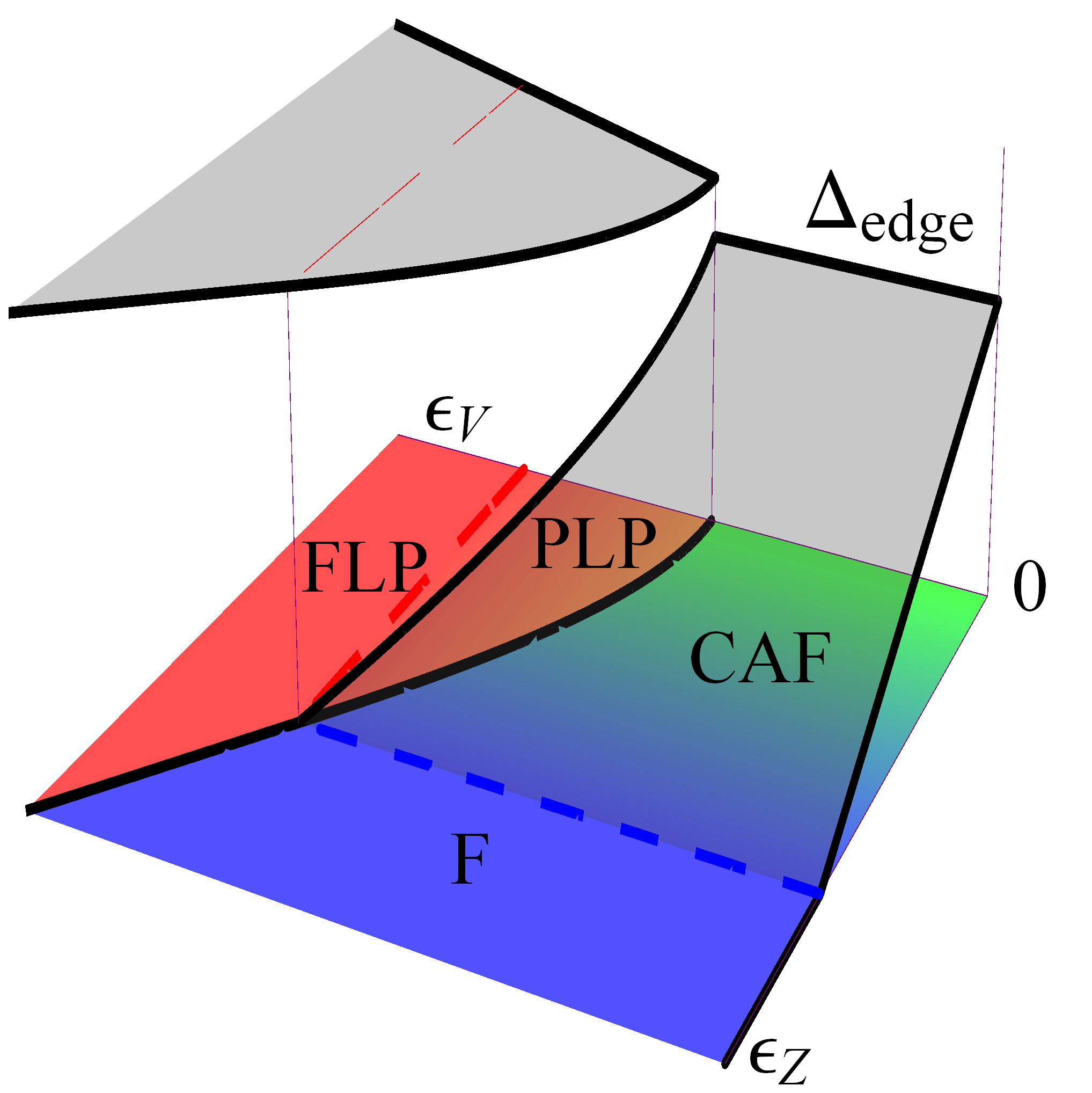}}
\caption{(color online). (left) Phase diagram in the space $(\e_V,\e_Z)$ of electric and Zeeman energies at fixed anisotropy energies $u_{\perp,z}$, for the case (\ref{eq:uAF})
of AF phase favored by the isospin anisotropy.
The violet line denotes the evolution of the system with applied electric field, realized in Ref.~\cite{h,Velasco},
with the CAF-PLP insulator-insulator transition at $\e_V=\e_V^*$ characterized by a conductance spike.
(right) Qualitative behavior of the edge transport gap $\De_\text{edge}$ as a function of $\e_V$ and $\e_Z$.
In the CAF phase, $\De_\text{edge}$ gradually decreases upon tilting the magnetic field and closes completely
once the F phase is reached.}
\label{fig:PDVZ}
\end{figure}

In real BLG,
the actual signs and ratio of
$u_{\perp,z}(B_\perp)$, which define the ground-state order at $\e_V=\e_Z=0$,
are determined
by details of e-e and e-ph interactions at the lattice scale.
Therefore, in practice, transitions between different phases can potentially be realized by varying the electric $\e_V$
or Zeeman $\e_Z=\mu_B B$ energies relative to the anisotropy energies $u_{\perp,z}(B_\perp)$ (Fig.~\ref{fig:PDVZ})
where the latter is achieved by tilting the magnetic field.

{\em Edge transport.}
Given the formal equivalence of the low-energy theories,
one can expect the edge charge excitations of the $\nu=0$ QHFM in BLG and MLG to have
qualitatively the same properties,
despite the differences in microscopic structures~\cite{BLGedge} of the edges.
Below we combine the earlier predictions for MLG~\cite{Abanin,FB,JM,Gusyninedge} with general physical arguments
to arrive at the anticipated~\cite{MKprep} edge transport phase diagram.

The charge of collective Skyrmion-type excitations
of the $\nu=0$ QHFM in BLG, associated with inhomogeneous isospin-spin textures $P(\rb)$,
is $2e$-quantized~\cite{APS} due to binding of electrons into pseudospin-singlet pairs.
Since in MLG {\em collective} edge excitations of the F phase are gapless~\cite{FB},
we conclude that the F phase in BLG supports gapless collective $2e$-charge edge excitations
and an ideal sample has a metallic $2e^2/h$
conductance {\em per edge}.

The AF and CDW phases in MLG are predicted to have gapped edge excitations~\cite{JM,Gusyninedge},
which can be seen as special cases of a more general property.
Noting that the isospin-singlet F phase is the only phase that does not break the valley symmetry,
one can argue that, in fact, all other orders of the $\nu=0$ QHFM
have gapped~\cite{edgegap} edge excitations. In particular,
the remaining CAF, PLP, and FLP phases are expected to be fully insulating
(note that bulk charge excitations are gapped in any phase).

Further distinction between the insulating phases is made by noticing the following properties.
The PLP ($0 < n_z^*< 1 $) and CAF ($0 < s_z^* < 1$) phases
continuously interpolate between their limiting cases,
ILC ($n_z^*=0$), FLP ($n_z^*=1$) and AF ($s_z^*=0$), F ($s_z^*=1$) phases, which can be tuned
by applying the electric field  ($\e_V$) and tilting the magnetic field ($\e_Z$),
respectively.
Therefore, CAF-F and PLP-FLP
are continuous second-order (at zero temperature) phase transitions (dashed blue and red lines in Figs.~\ref{fig:PD} and \ref{fig:PDVZ})
and no sudden changes in transport properties, such as a conductance spike, should occur there.
Consequently, first, the system remains insulating
as it transitions from the PLP to FLP phase, without a sharp feature of the PLP-FLP transition in transport.
Second, since AF and F phases have gapped
and gapless edge charge excitations, respectively,
we are forced to conclude, by continuity,
that the edge transport gap $\De_\text{edge}^\text{CAF}(s_z^*)$
of the CAF phase monotonically decreases with $s_z^* =  \e_Z/(2 |u_\perp|)$
from a finite value $\De_\text{edge}^\text{CAF}(s_z^*\!=\!0) = \De_\text{edge}^\text{AF}$ at $\e_Z=0$
to zero $\De_\text{edge}^\text{CAF}(s_z^*\!=\!1) \!=\! \De_\text{edge}^\text{F} \!=\!0$ at the CAF-F boundary $\e_Z \!=\! 2 |u_\perp|$.
I.e., the continuous transformation of the CAF to F phase
upon tilting the field is accompanied by gradual closing of the edge transport gap.
In contrast to the latter, the edge transport gap of the {\em spin-singlet} PLP and FLP phases
is not expected to appreciably depend on $\e_Z$.

On the other hand, CAF-PLP and F-FLP are discontinuous first-order phase transitions (black solid lines in Figs.~\ref{fig:PD} and \ref{fig:PDVZ}),
which could be signified by conductance spikes due to increased symmetry at the transition lines.
The FLP-F is an insulator-metal transition,
whereas the CAF-PLP is an insulator-insulator transition.
The resulting qualitative dependence of the edge transport gap $\De_\text{edge}$
is plotted in Fig~\ref{fig:PDVZ}.

{\em Canted antiferromagnetic phase.}
We now identify
the  insulating $\nu=0$ phase observed in Refs.~\cite{h,Velasco} at electric energy
$\e_V<\e_V^*$ below the critical value $\e_V^* \approx E^* a_z/2 \approx 20 B_\perp [\Ttxt] \Ktxt$.
The F phase at $\e_V<\e_V^*$ is ruled out as having metallic $2e^2/h$ edge conductance.
The phase at high enough
$\e_V\gtrsim \e_V^*$
is readily identified as the FLP phase and hence it cannot also occur at lower electric field.
The PLP phase at $\e_V < \e_V^*$ (ILC at $\e_V=0$) is ruled out, since
the transition between the PLP and FLP phases is continuous
and the system would be insulating at all $\e_V$ values.
The phase at $\e_V<\e_V^*$
is therefore the remaining insulating CAF phase of the $\nu=0$ QHFM.
The evolution of the system with applied electric field is denoted by a violet line in Fig.~\ref{fig:PDVZ}.
The conductance spike at $\e_V=\e_V^*$ thus corresponds to the CAF-PLP
insulator-insulator transition, which is the only such transition on the phase diagram (Figs.~\ref{fig:PD} and \ref{fig:PDVZ});
upon further increasing the electric field, the PLP phase continuously transitions to the FLP phase,
with the system remaining insulating at $\e_V>\e_V^*$.

The conclusion about the CAF phase implies
that the isospin anisotropy (\ref{eq:Edm})
favors the AF  phase, i.e., that in real BLG the case
\beq
    u_z> -u_\perp >0
\label{eq:uAF}
\eeq
is realized. This is consistent with microscopic considerations.
In BLG, the leading anisotropy $u_z>0$ arises from e-e interactions due to a finite layer separation.
This ``capacitance effect'' favors equal charge population of the
layers and the anisotropy energy $\Ec_\dm(P)$ with only $u_z > 0$ present
is minimized by the states $\chi_a=|K\ran \otimes |\s_a\ran$, $\chi_b=|K'\ran \otimes |\s_b\ran$
with {\em arbitrary} spin polarizations $\s_{a,b}$ of the layers
(it can also be demonstrated~\cite{MKprep} that at this level these are, in fact, {\em exact eigenstates} at any layer separation,
which suggests their particular robustness).
This spin degeneracy is then lifted by the competition between the anisotropy $u_\perp$ and the Zeeman effect.
The negative $u_\perp<0$ favoring antiferromagnetic order can naturally arise from e-ph or renormalized e-e interactions~\cite{eph,MKmono,CAFbilayer}.

The critical electric energy $\e_V^*$ is related to $u_{\perp,z}$ and $\e_Z$ according to Eq.~(\ref{eq:CAFPLP}) for the CAF-PLP transition.
Given the large discrepancy between $\e_V^*$ and $\e_Z \approx 0.7 B[\Ttxt] \Ktxt$ at moderate tilt angles, one may neglect $\e_Z$ to obtain
\beq
    \e_V^*  =  \sqrt{u_z^2-u_\perp^2}.
\label{eq:eV*f}
\eeq
At smaller $|u_\perp|$, $\e_V^*$ is mainly determined by $u_z$:
at $|u_\perp| \lesssim u_z/2$, one  may also neglect $|u_\perp|$ with decent accuracy to extract
the anisotropy $u_z \approx \e_V^* \approx 20 B_\perp[\Ttxt]\Ktxt$.
The magnitude and linear $B_\perp$-dependence of $\e_V^*$ at higher $B_\perp \gtrsim 2\Ttxt$ in Ref.~\cite{h}
are fully consistent with
the properties  of the bare anisotropies $ u^{(0)}_{\perp,z}(B_\perp)$,
whereas the deviation from linearity at lower $B_\perp \lesssim 2\Ttxt$  can be
explained by enhanced renormalizations of $u_{\perp,z}(B_\perp)$
as the transition to the low-magnetic-field interaction-induced state of debated nature~\cite{Castro,NL,Zhang,VY,LATF} is approached.

{\em Tilted-field experiment.}
Tilting the magnetic field by a moderate $45^\circ$ angle ($B/B_\perp=\sqrt{2}$) in Ref.~\cite{h}
resulted in a small yet systematic increase of the critical $\e_V^*$ value,
which is also consistent with Fig.~\ref{fig:PDVZ} and Eq.~(\ref{eq:CAFPLP}),
but did not induce any new phase transitions.
However, according to
the phase diagram in Fig.~\ref{fig:PDVZ},
upon further increasing the tilt ratio $B/B_\perp$,
the  CAF-F and FLP-F transitions  will eventually occur.
The tilt ratio
$
    B/B_\perp = 2 |u_\perp|/(\mu_B B_\perp)
$
required for reaching the F phase is determined by the value of $|u_\perp|$,
which cannot be obtained from $\e_V^*$ [Eq.~(\ref{eq:eV*f})] independently of $u_z$
and for smaller $|u_\perp|\lesssim u_z/2$ remains essentially unknown.
The most favorable case would be $|u_\perp| \ll u_z$.
For reference, at ``large'' $|u_\perp|= u_z/2 \approx 10 B_\perp[\Ttxt]\Ktxt$,
the required tilt ratio is
$B/B_\perp \approx 30$.
Since the low-electric-field insulating $\nu=0$ state
is detectable at as low as $B_\perp \approx 1 \Ttxt$~\cite{h,SchweizBLG,Velasco},
the practical tilt ratio as high as $B/B_\perp \sim 50$ could be achieved in BLG
for available static magnetic fields $B \leq 45 \Ttxt$.

{\em Outlook.}
Provided the F phase can be reached
by tilting the magnetic field,
it becomes possible to explore the whole phase diagram (Fig.~\ref{fig:PDVZ}) of the $\nu=0$ QHFM in dual-gated BLG devices.
The predicted marked distinction
between the edge transport properties of the ``spin-active'' CAF and spin-singlet FLP phases
-- gradual closing of the edge gap $\De_\text{edge}$ with tilting the field vs its insensitivity --
should manifest itself in such an experiment
as continuous CAF-F vs sharp FLP-F insulator-metal transitions.
These features can also be used to test the presented theory and distinguish between the phases in the experiment.

Author is thankful to P. Coleman for insightful discussions and helpful comments on the manuscript
and to E. Andrei, C. N. Lau, A. Young, and M. Foster for insightful discussions.
The work was supported by the U.S. DOE Grants No.~DE-FG02-99ER45790 and No.~DE-AC02-06CH11357.


\begin{thebibliography}{99}
\bibitem{biasfield}
Note that $E$ is indeed the field created by the gates,
while the ``self-action'' layer charge imbalance effects are systematically taken into account by the anisotropy (\ref{eq:Edm}).
\bibitem{QFs}
Note that quantum fluctuations do not diverge in the infrared in $2+1=3$ space-time dimensions
and are therefore not essential for QHFM systems.
\bibitem{edgegap}
Due to strong screening~\cite{GorbarBLG},
bulk and edge transport gaps in BLG scale as
$\hbar^2/(m l_B^2)$ ($m$ is the effective mass).
\end{thebibliography}
\end{document}